\documentclass[12pt,a4paper]{article}

\usepackage{amsmath}
\usepackage{graphicx}

\begin{document}

\title{Electric charge in hyperbolic motion: arcane geometrical aspects}
\author{C\u alin Galeriu}

\maketitle

\section*{Abstract}

{\it We look at an electric charge in hyperbolic motion, 
and we describe some arcane geometrical aspects of the electrodynamic interaction. 
We discuss the advantages of a time symmetric formulation in which the material point particles are 
replaced by infinitesimal length elements along their worldlines. We show that the four-force obtained 
from our geometrical model is very closely related to the "electrostatic" four-force obtained 
from Fokker's time symmetric action.}

\section{Introduction}

The study of an electric charge in hyperbolic motion is an important aspect of Minkowski's geometrical formulation of electrodynamics. 
In "Space and Time" \cite{minkowski}, his last publication before his premature death, Minkowski gave a brief geometrical recipe for calculating the four-force 
with which an electric charge acts on another electric charge. 
In a previous article \cite{galeriuHYP} we have discussed Minkowski's geometrical recipe and have rederived his expression of the four-force, 
in an effort to provide a more modern, accessible, 
and unified presentation of the early history of the electric charge in hyperbolic motion. This second article continues the study of the electric charge
in hyperbolic motion, and a thorough study of our previous article \cite{galeriuHYP} is necessary. Here we focus on some highly speculative 
observations, all of them based on some arcane geometrical aspects of the electrodynamic interaction. 
These geometrical relations are discovered with the help of Minkowski space diagrams.

\begin{figure}[h!]
\begin{center}
\includegraphics[height=9.0cm]{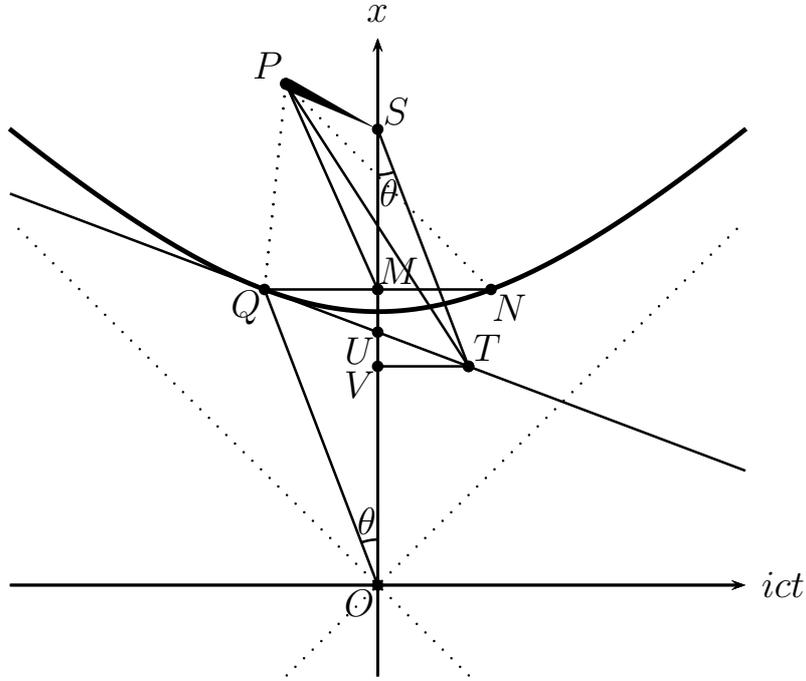}
\caption{In the co-moving reference frame of the source particle at $Q$, the test particle at $P$ is described by a radial position vector $TP$, of length $r$.
In the reference frame in which the test particle at $P$ and the center of the hyperbola $O$ are simultaneous the electric field has the direction of vector $MP$
and the magnetic field is zero.}
\label{fig:combined}
\end{center}
\end{figure}

We start by combining Figures 3 and 4 of \cite{galeriuHYP} into just one, shown here as Figure \ref{fig:combined}. It was shown in \cite{galeriuHYP}
that the electric field produced at point $P$ by the electric charge $Q$ in hyperbolic motion has the magnitude $Q/r^2$ (in Gaussian units) and the 
direction of the vector $\overrightarrow{MP}$.

\section{The Geometrical Argument Supporting Time Symmetric Electrodynamics}

We have seen that the only effect that the acceleration of the source particle has on the electric field produced at the position of the test particle 
is a change in direction, while the magnitude of the electric field stays the same. What is the process by which the electric field ends up with 
the direction of segment $\overrightarrow{MP}$? 
Point $P$ is of course selected as the point where the electric field acts, but what is the mechanism by which point $M$ is selected? 

A careful look at Figure \ref{fig:combined} shows that point $M$ is the midpoint of segment $QN$, where $Q$ is the position of the retarded source charge 
and $N$ is the position of the advanced source charge. We argue that point $M$ defines the direction of the electric field because the electrodynamic 
interaction is time symmetric. Instead of decomposing the segment $\overrightarrow{MP}$ as in \cite{galeriuHYP}:
\begin{equation}
\overrightarrow{MP} = \overrightarrow{QP} - \overrightarrow{QM},
\label{eq:sommerfeld}
\end{equation}
we notice that we can also decompose the segment $\overrightarrow{MP}$ as:
\begin{equation}
\overrightarrow{MP} = \frac{1}{2} (\overrightarrow{QP} + \overrightarrow{NP}).
\label{eq:galeriu}
\end{equation}

This geometrical description of the electric field is so amazingly simple and beautiful, it brings to life Minkowski's vision: 
"[...] physical laws might find their most perfect expression as reciprocal relations between these worldlines." \cite{minkowski} 
The time symmetric nature of this electromagnetic field, produced by an electric charge in hyperbolic motion in the reference frame of Figure \ref{fig:combined}, 
is very easy to understand. It was already noticed by Boulware that "[...] under time reversal, retarded fields are transformed into advanced fields, 
hence the retarded field is equal to the advanced field." \cite{boulware} This result will hold true even after a Lorentz transformation.
The same time symmetric geometrical description applies to the case of a source charge in uniform motion \cite{galeriuAAAD}. 
When the source charge moves with constant velocity, or with constant acceleration, our geometrical description is identical to the generally accepted 
causal theory. Since most of classical electrodynamics is concerned with such motion of the particles, one could hardly find any experimental evidence 
against a time symmetric formulation. Even if the classical electric charges had a variable acceleration, 
that would not be reflected in the electromagnetic field produced. 
The variable acceleration does shows up in the expression of the radiative damping force, but that is still a controversial topic \cite{galeriuRR}. 
In classical electrodynamics the instantaneous radiation reaction force is outside of direct experimental investigation.

The electric field $\overrightarrow{E} = (E_x, E_y, E_z, 0)$ defined by (\ref{eq:galeriu}) has two contributions, one retarded and one advanced. 
\begin{equation}
\overrightarrow{E} = \frac{Q}{r^2} \frac{\overrightarrow{MP}}{MP} = \frac{Q}{2 r^2} (\frac{\overrightarrow{QP}}{MP} + \frac{\overrightarrow{NP}}{MP}).
\label{eq:1}
\end{equation}
The three components of the 
electric field do not make a four-vector in general, but one can look at $q (E_x, E_y, E_z, 0)$
as the four-force that acts on a particle with electric charge $q$ and at rest in the unique
reference frame of Figure \ref{fig:combined}. This four-force has two contributions, one retarded $F_{QP}$ and one advanced $F_{NP}$, which 
look like contact forces in Minkowski space, since $QP = NP = 0$.

Let's look at the retarded component along $\overrightarrow{QP}$. Since $\overrightarrow{QP} = \overrightarrow{QM} + \overrightarrow{MP}$, 
and $QM = i\ MP$,
we expect an electric four-force component of magnitude $F_{QM} = i \frac{1}{2} \frac{Q q}{r^2}$ along $\overrightarrow{QM}$
and an electric four-force component of magnitude $F_{MP} = \frac{1}{2} \frac{Q q}{r^2}$ along $\overrightarrow{MP}$. 
\begin{equation}
\overrightarrow{F_{QP}} = \frac{Q q}{2 r^2} \frac{\overrightarrow{QP}}{MP} 
= \frac{Q q}{2 r^2} \frac{\overrightarrow{QM}}{MP} + \frac{Q q}{2 r^2} \frac{\overrightarrow{MP}}{MP}
= \frac{i Q q}{2 r^2} \frac{\overrightarrow{QM}}{QM} + \frac{Q q}{2 r^2} \frac{\overrightarrow{MP}}{MP}
= \overrightarrow{F_{QM}} + \overrightarrow{F_{MP}} 
\label{eq:2}
\end{equation}
A similar retarded contribution can be evaluated 
in the reference frame that is co-moving with the source charge at $Q$. Since $\overrightarrow{QP} = \overrightarrow{QT} + \overrightarrow{TP}$, 
and $QT = i\ TP$,
we should be able to calculate the electric four-force component $F_{TP}$ along $\overrightarrow{TP}$  and relate it to the 
electric four-force component $F_{MP}$ along $\overrightarrow{MP}$. 
\begin{equation}
\overrightarrow{F_{QP}} = \overrightarrow{F_{QT}} + \overrightarrow{F_{TP}} = F_{QT} \frac{\overrightarrow{QT}}{QT} + F_{TP} \frac{\overrightarrow{TP}}{TP}
\label{eq:3}
\end{equation}
Let $V$ be the projection of point $T$ on the $Ox$ axis. Since $\overrightarrow{TP} = \overrightarrow{TV} + \overrightarrow{VM} + \overrightarrow{MP}$, 
it follows that
\begin{equation}
\overrightarrow{F_{TP}} = F_{TP} \frac{\overrightarrow{TP}}{TP} = F_{TP} \left( \frac{\overrightarrow{TV}}{TP} 
+ \frac{\overrightarrow{VM}}{TP} + \frac{\overrightarrow{MP}}{TP} \right).
\label{eq:4}
\end{equation}
Substituting (\ref{eq:4}) into (\ref{eq:3}) we have
\begin{equation}
\overrightarrow{F_{QP}} = F_{QT} \frac{\overrightarrow{QT}}{QT} + F_{TP} \left( \frac{\overrightarrow{TV}}{TP} 
+ \frac{\overrightarrow{VM}}{TP} + \frac{\overrightarrow{MP}}{TP} \right).
\label{eq:5}
\end{equation}
From (\ref{eq:2}) and (\ref{eq:5}) we can write
\begin{equation}
\frac{i Q q}{2 r^2} \frac{\overrightarrow{QM}}{QM} + \frac{Q q}{2 r^2} \frac{\overrightarrow{MP}}{MP} 
= F_{QT} \frac{\overrightarrow{QT}}{QT} + F_{TP} \left( \frac{\overrightarrow{TV}}{TP} 
+ \frac{\overrightarrow{VM}}{TP} + \frac{\overrightarrow{MP}}{TP} \right),
\label{eq:6}
\end{equation}
and since $\overrightarrow{QM}$, $\overrightarrow{QT}$, $\overrightarrow{TV}$, and $\overrightarrow{VM}$ 
are restricted to the $(x,O,ict)$ plane, it follows that 
\begin{equation}
F_{TP} = \frac{Q q}{2 r^2} \frac{TP}{MP}. 
\label{eq:7}
\end{equation}
Since $TP = r = - i \rho \sin(\theta)$ \cite{galeriuHYP} (p. 369), $QM = a \sin(\theta)$ \cite{galeriuHYP} (p. 371), $QM = i\ MP$, 
and $MP = QM/i = - i a \sin(\theta)$, it follows that
\begin{equation}
F_{TP} = \frac{Q q}{2 r^2} \frac{\rho}{a}. 
\label{eq:8}
\end{equation}

\section{The Geometrical Argument Against the Material Point Particle Model}

We notice that, if the electric source charge $Q$ at point $Q$ had a constant velocity, moving along the worldline $QT$,
the retarded contribution to the electric field at point $P$ would be just $Q/(2 r^2)$. The retarded contribution to the four-force
acting on a electric test charge $q$ at rest relative to the source charge would be just $Qq/(2 r^2)$. 
One has to recognize that a test charge at $P$, at rest in the reference frame
in which the point $P$ and the center of the hyperbola $O$ are simultaneous, is not the same as a test charge at $P$, at rest 
in the reference frame co-moving with the source particle at $Q$. The state of motion of the test particle is essential 
to the determination of the four-force acting on the test particle.
What is the geometrical explanation of the factor $\rho/a$ introduced into the expression of the four-force by a change in the state of
motion of the test particle?

In other words, suppose that we have a test charge $q$ in an electric field $\vec{E}$, with no magnetic field present. 
The force acting on the test charge is $\vec{F} = q \vec{E}$. If the velocity $\vec{v}$ of the particle is zero, 
then its four-velocity is $(\vec{0}, i c)$ and the four-force is $(\vec{F}, 0)$. But if the velocity $\vec{v}$ is not zero, 
then its four-velocity is $(\gamma \vec{v}, i \gamma c)$ and the four-force is $(\gamma \vec{F}, i \frac{\gamma}{c} \vec{F} \cdot \vec{v} )$,
where $\gamma = 1 / \sqrt{1 - \frac{v^2}{c^2}}$. 
Why is the four-force dependent on the velocity of the test charge?

In a similar context \cite{galeriuAAAD}, when looking at the interaction between two electrically charged particles in uniform motion, we 
have noticed that the four-force shows an explicit dependence on the velocity of the test particle, and an implicit dependence on the 
velocity of the source particle (through the chosen reference frame). However, "from a {\it geometrical} point of view, a point in Minkowski space
is just a fixed point - it does not have a velocity!" \cite{galeriuAAAD} The conclusion was that,
from a geometrical point of view, the electrodynamic four-force, the relevant 
Minkowski vector describing the interaction, is incompatible with the material point particle model.

The solution we proposed \cite{galeriuAAAD} was to recognize the similarity between the worldline of a particle and a string in static equilibrium,
and to recognize that instead of a four-force acting on a point particle we actually have a four-force linear density 
acting on an infinitesimal length element along
the particle's worldline. The electrodynamic interaction takes place between infinitesimal worldline segments with a very special property: 
they have their start and end points connected by light signals.
The ratio of the length of these segments changes with the velocity of the particles, thus explaining the velocity dependence of the four-force. 
We also need to have a time symmetric action-at-a-distance electrodynamic interaction,
in order to obtain some agreement with the classical expression of the four-force.

Another interesting observation is that one cannot define a relativistically invariant distance starting with just two spacetime points 
connected by a light signal \cite{galeriuAAAD}. 
The distance that enters Coulomb's law is measured in the reference frame in which the source particle is at rest. To determine this 
reference frame we need at least two infinitesimally close points on the worldline of the source particle. 
Therefore, the point particle model has to be replaced with a new model in which the particles are segments of infinitesimal (or very small) length
along their worldlines. In a different context, exploring the concept of inertia, 
Kevin Brown has reached a similar conclusion: "[...] even an object with zero spatial extent has non-zero temporal extent [...]". \cite{brown} (page 169)

Several other authors have experienced a similar insight, recognizing the inadequacy of the material point particle model. 
Colin Lamont draws a light cone with a variable thickness, in order to understand why the denominator of the four-potential has directional dependence, and writes:
"A helpful image is that of a lightcone as a region of infinitesimal thickness bounded by the level sets $r^\mu r_\mu - \epsilon = 0$ and $r^\mu r_\mu + \epsilon = 0$ for some infinitesimal positive parameter $\epsilon$. Then the strength of the potential produced by a charge at this point is determined by how much proper time the charge spends in this region." \cite{lamont}
Kevin Brown draws a light cone with a constant thickness, in order to understand how the electric field transforms from one reference frame to another, and writes: "The light cone is shown with a non-zero thickness to illustrate that the duration of time spent by each particle as it passes through the light cone depends on the speed of the particle. The basic charge $q$ is defined based on the static Coulomb potential, represented by the vertical ($v = 0$) path. In general, the duration of coordinate time spent by a point-like particle in the light cone shell is proportional to $1/(1 + v/c)$. This proportionality remains the same, regardless of the thickness of the light cone shell, even in the limit as the thickness goes to zero." \cite{brown} (page 228)

After the publication of \cite{galeriuAAAD} we have discovered that Costa de Beauregard \cite{costa} had also noticed, a long time ago, the similarity 
between the worldline of a particle 
and a string in static equilibrium. In spite of this, Costa de Beauregard has continued to use the venerable material point particle model,
together with its associated Dirac delta functions, in all of his physics books and articles. We will now briefly describe the reason why a worldline 
is similar to a string in static equilibrium. A point particle with four-momentum $\mathbf{P}$, subject to a four-force $\mathbf{F}$, undergoes
a trajectory described by 
\begin{equation}
\mathbf{P}_B - \mathbf{P}_A = \mathbf{F} d\tau,
\label{eq:worldline}
\end{equation}
where $A$ and $B$ are two infinitesimally close points on the worldline of the particle, and $\tau$ is the proper time.
A static string with tension $\mathbf{T}$, subject to a linear force density $\mathbf{f}$, has all of its segments of infinitesimal length $ds$ in static equilibrium.
This static equilibrium condition takes the form 
\begin{equation}
\mathbf{T}_B + (- \mathbf{T}_A) + \mathbf{f} ds = 0, 
\label{eq:string}
\end{equation}
where the minus sign indicates that the tension forces
acting at the end points of the segment $ds$ have different directions. The equations (\ref{eq:worldline}) and (\ref{eq:string}) are equivalent provided that 
$ds = i\ c\ d\tau$, $\mathbf{T} = i\ c\ \mathbf{P} / s_o$, and $\mathbf{f} = - \mathbf{F} / s_o$. 
Here $s_o = i\ c\ \tau_o$ is a very small constant quantity with units of length, which we call
"the length of a particle". We also notice that the four-force acting on the material point particle and the four-force density acting on the
infinitesimal worldline segment have opposite directions. We therefore have to consistently swap attraction and repulsion between forces 
and linear force densities when going from
one model to the other. 

In our model we imagine the worldline of a particle as a string in Minkowski space, under a tension and in static equilibrium. 
While the generally accepted view is that the magnitude of the
four-momentum of a particle is constant (the rest mass is constant), and as a result the magnitude of the tension in the 
worldline string is constant, this is
not an absolutely necessary requirement. It was already noticed by Tetrode \cite{tetrode} a long time ago that in the most general case of time symmetric
electrodynamic interactions the rest masses of the particles may depend upon their four-accelerations. In this most general case, 
we have to write $\int m\ ds$ instead of $m \int ds$ in the expression of the action.
We have also pointed out that, in the case of motion with variable acceleration, the four-force obtained from our model \cite{galeriuAAAD} is not always orthogonal 
to the four-velocity. A small deviation from orthogonality could indeed exist, and our geometrical recipe in this case amounts to supplementing the antisymmetric electromagnetic field tensor $\mathcal{F}_{ik}$ with a small tensor $\mathcal{G}_{ik}$. We expect the contribution of $\mathcal{G}_{ik}$ to average out to 
zero, and we were able to prove this conjecture in a very simple situation \cite{galeriuAAAD}. 
In any case, by requiring that the rest mass of an electron at a given point in spacetime does not depend on its history, 
and by using Stoke's theorem, we have obtained the two homogeneous Maxwell equations \cite{galeriuMax}, which in itself is a remarkable result.

In our previous work \cite{galeriuAAAD} we have kept the length of the test particle constant, 
while allowing for the length of the source particle to change.
This time the length of the source particle will be held constant, 
while allowing for the length of the test particle to change. This new approach is, on a procedural level, 
in better agreement with the derivation of the equation of motion of the test particle
from a variational principle. Indeed, in the variational method \cite{fokker} the worldlines of the source particles are held constant, 
while the worldline of the test
particle is allowed to have a small variation. 

\section{One Dimensional Motion}

\subsection{Source Charge at Rest}

\begin{figure}[h!]
\begin{center}
\includegraphics[height=77pt]{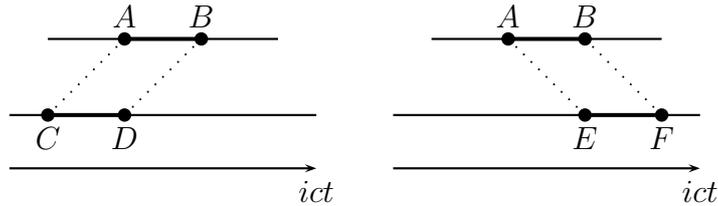}
\caption{A test particle $AB$, at rest relative to the source particle at rest, feels a retarded interaction from $CD$ and an advanced interaction from $EF$.}
\label{fig:atrest}
\end{center}
\end{figure}

The first situation, shown in Figure \ref{fig:atrest}, has the source charge $Q$ at rest at the origin, and the test charge $q$ also at rest. 
The distance, along the $x$ axis, between the two particles is $R$. 
The four-force linear density has a retarded part along $CA$, with components 
\begin{equation}
\frac{-1}{s_o}(\frac{Q q}{2 R^2}, 0, 0, i \frac{Q q}{2 R^2}),
\end{equation}
and an advanced part along $EA$, with components 
\begin{equation}
\frac{-1}{s_o}(\frac{Q q}{2 R^2}, 0, 0, - i \frac{Q q}{2 R^2}).
\end{equation} 
The net four-force linear density has the components 
\begin{equation}
\frac{-1}{s_o}(\frac{Q q}{R^2}, 0, 0, 0),
\end{equation} 
and the four-force has the components
\begin{equation}
(\frac{Q q}{R^2}, 0, 0, 0).
\end{equation}

\begin{figure}[h!]
\begin{center}
\includegraphics[height=105pt]{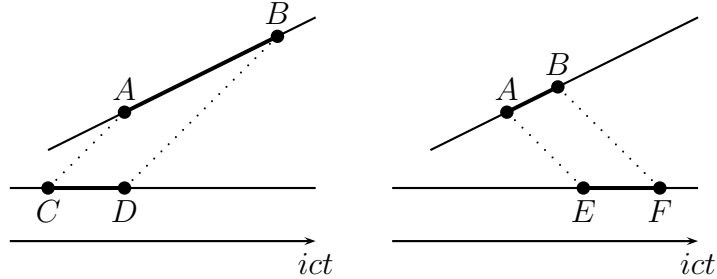}
\caption{A test particle $AB$, in motion relative to the source particle at rest, feels a retarded interaction from $CD$ and an advanced interaction from $EF$.}
\label{fig:inmotion}
\end{center}
\end{figure}

The second situation, shown in Figure \ref{fig:inmotion}, has the source charge $Q$ at rest at the origin, 
and the test charge $q$ moving with a radial velocity $v$. 
The distance, along the $x$ axis, between the two particles is $R$. 
The ratios of the segments are 
$\frac{AB}{CD} = \frac{\sqrt{1-\frac{v^2}{c^2}}}{1-\frac{v}{c}}$ (retarded interaction) and
$\frac{AB}{EF} = \frac{\sqrt{1-\frac{v^2}{c^2}}}{1+\frac{v}{c}}$ (advanced interaction) \cite{galeriuAAAD}.
The four-force linear density has a retarded part along $CA$, with components 
\begin{equation}
\frac{-1}{s_o}(\frac{Q q}{2 R^2}\frac{\sqrt{1-\frac{v^2}{c^2}}}{1-\frac{v}{c}}, 0, 0, i \frac{Q q}{2 R^2}\frac{\sqrt{1-\frac{v^2}{c^2}}}{1-\frac{v}{c}}),
\end{equation}
and an advanced part along $EA$, with components 
\begin{equation}
\frac{-1}{s_o}(\frac{Q q}{2 R^2}\frac{\sqrt{1-\frac{v^2}{c^2}}}{1+\frac{v}{c}}, 0, 0, - i \frac{Q q}{2 R^2}\frac{\sqrt{1-\frac{v^2}{c^2}}}{1+\frac{v}{c}}).
\end{equation}
The net four-force linear density has the components
\begin{equation}
\frac{-1}{s_o}(\frac{Q q}{R^2}\frac{1}{\sqrt{1-\frac{v^2}{c^2}}}, 0, 0, i \frac{Q q}{R^2}\frac{v}{c}\frac{1}{\sqrt{1-\frac{v^2}{c^2}}}),
\end{equation}
and the four-force has the components
\begin{equation}
(\frac{Q q}{R^2}\frac{1}{\sqrt{1-\frac{v^2}{c^2}}}, 0, 0, i \frac{Q q}{R^2}\frac{v}{c}\frac{1}{\sqrt{1-\frac{v^2}{c^2}}}).
\end{equation}

\subsection{Source Charge in Hyperbolic Motion}

\begin{figure}[h!]
\begin{center}
\includegraphics[height=212pt]{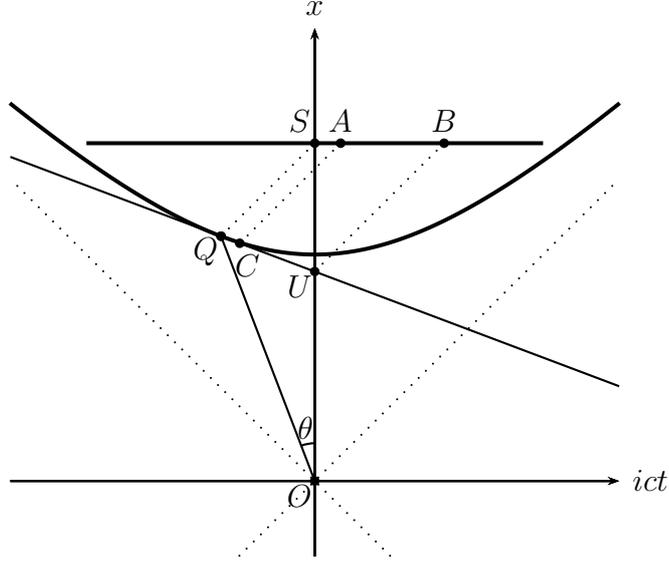}
\caption{A test particle $SA$, at rest in a reference frame in which the source particle $QC$ is in hyperbolic motion, 
and in which the test particle and the center of the hyperbola $O$ are simultaneous, feels a retarded interaction from $QC$. 
Due to the time symmetry of this situation, the advanced interaction is caused by an infinitesimal segment of length equal to $QC$, not shown in this figure.}
\label{fig:hyperbola}
\end{center}
\end{figure}

The third situation, shown in Figure \ref{fig:hyperbola}, has the source charge in hyperbolic motion, while the test charge is at rest and simultaneous
with the center of the hyperbola. The source particle, the infinitesimal segment $QC$, interacts with the test particle, the infinitesimal segment $SA$.
The ratio of these segments is $\frac{SA}{QC} = \frac{SB}{QU}$. Since $US = \rho - \frac{a}{\cos(\theta)}$ \cite{galeriuHYP} (p. 368-369), it follows that 
$SB = i US = i \left( \rho - \frac{a}{\cos(\theta)} \right)$. Since $\frac{QU}{OQ} = \tan(\theta)$ and $OQ = a$, it follows that $QU = a \tan(\theta)$.
As a result $\frac{SB}{QU} = \frac{i \left( \rho - \frac{a}{\cos(\theta)} \right)}{a \frac{\sin(\theta)}{\cos(\theta)}} 
= \frac{\rho \cos(\theta) - a}{- i a \sin(\theta)} = \frac{\rho}{a} \frac{\rho \cos(\theta) - a}{- i \rho \sin(\theta)}$.
When $y = z = 0$ we have $r = \rho \cos(\theta) - a$ \cite{galeriuHYP} (p. 369), and since $r = - i \rho \sin(\theta)$, it follows that in this example of 
one dimensional motion the ratio of the infinitesimal segments is $\frac{SA}{QC} = \frac{\rho}{a}$. This is in agreement with (\ref{eq:8}).
In the reference frame of Figure \ref{fig:hyperbola} 
the four-force linear density has a retarded part with components 
\begin{equation}
\frac{-1}{s_o}(\frac{Q q}{2 r^2}, 0, 0, i \frac{Q q}{2 r^2}),
\end{equation}
and an advanced part with components 
\begin{equation}
\frac{-1}{s_o}(\frac{Q q}{2 r^2}, 0, 0, - i \frac{Q q}{2 r^2}). 
\end{equation}
The net four-force linear density has the components 
\begin{equation}
\frac{-1}{s_o}(\frac{Q q}{r^2}, 0, 0, 0),
\end{equation}
and the four-force has the components
\begin{equation}
(\frac{Q q}{r^2}, 0, 0, 0).
\end{equation}

The fourth situation has the source charge in hyperbolic motion, while the test charge is simultaneous
with the center of the hyperbola, like before. However, this time the test particle is moving along the $x$ axis with a constant radial velocity $v$.
Since the retarded and advanced infinitesimal segments of the source particles on the hyperbolic worldline are the same as in the previous case, and
the lightcones from the start and end points of the infinitesimal segments of the source particles are
the same, the intersection of these lightcones with the worldline of the test particle produces segments identical
to those in Figure \ref{fig:inmotion}. We understand in this way why, in this special reference frame, 
the four-force depends on the radial velocity of the particle.
In the reference frame of Figure \ref{fig:hyperbola} 
the linear four-force density has a retarded part with components 
\begin{equation}
\frac{-1}{s_o}(\frac{Q q}{2 r^2}\frac{\sqrt{1-\frac{v^2}{c^2}}}{1-\frac{v}{c}}, 0, 0, i \frac{Q q}{2 r^2}\frac{\sqrt{1-\frac{v^2}{c^2}}}{1-\frac{v}{c}}),
\end{equation}
and an advanced part with components 
\begin{equation}
\frac{-1}{s_o}(\frac{Q q}{2 r^2}\frac{\sqrt{1-\frac{v^2}{c^2}}}{1+\frac{v}{c}}, 0, 0, - i \frac{Q q}{2 r^2}\frac{\sqrt{1-\frac{v^2}{c^2}}}{1+\frac{v}{c}}).
\end{equation}
The net linear four-force density has the components
\begin{equation}
\frac{-1}{s_o}(\frac{Q q}{r^2}\frac{1}{\sqrt{1-\frac{v^2}{c^2}}}, 0, 0, i \frac{Q q}{r^2}\frac{v}{c}\frac{1}{\sqrt{1-\frac{v^2}{c^2}}}),
\end{equation}
and the four-force has the components
\begin{equation}
(\frac{Q q}{r^2}\frac{1}{\sqrt{1-\frac{v^2}{c^2}}}, 0, 0, i \frac{Q q}{r^2}\frac{v}{c}\frac{1}{\sqrt{1-\frac{v^2}{c^2}}}),
\end{equation}

In conclusion, in the one dimensional case our geometrical model with time symmetric interactions reproduces correctly the 
textbook expressions of the four-force.

\subsection{The Law of Action and Reaction}

We will now focus our attention on the third situation described above, with the source charge in hyperbolic motion 
and the test particle at rest and simultaneous with the center of the hyperbola, to see how the law of action and reaction 
applies in the case of one dimensional motion. Since all that really matters is the source particle $QC$ and the test particle $SA$,
the situation analyzed is in fact more general, and does not really require the source charge to be in hyperbolic motion.

In the reference frame in which the source particle $QC$ is at rest, a test particle also at rest and at a distance $ST = r$ from the source 
will feel a retarded four-force density with a real component $\frac{-1}{s_o} \frac{Q q}{2 r^2}$ along $\overrightarrow{TS}$. But when the test particle is in
motion relative to the source particle, and at rest relative to the reference frame in which the test particle and 
the center of the hyperbola are simultaneous, the test particle will feel a retarded four-force density with a real component 
$\frac{-1}{s_o} \frac{Q q}{2 r^2} \frac{SA}{QC} = \frac{-1}{s_o} \frac{Q q}{2 r^2} \frac{\rho}{a}$ along $\overrightarrow{TS}$. The projection
of this four-force density along $\overrightarrow{OS}$ will be $\frac{-1}{s_o} \frac{Q q}{2 r^2}$, as discussed at the begining of this paper.
The real component of the retarded four-force with which particle $QC$ is acting on particle $SA$, along $\overrightarrow{OS}$, in the worldline string model, is 
\begin{equation}
\frac{-1}{s_o} \frac{Q q}{2 r^2} SA.
\label{eq:action}
\end{equation}
This expression (\ref{eq:action}) turns into the usual expression $\frac{Q q}{2 r^2}$ of the material point particle model 
when $SA$ is substituted with $s_o$ and attraction turns into repulsion, but for now we continue working with the worldline string model
and we do not impose any specific length on the source and test particles.
We want to find the advanced four-force with which particle $SA$ is acting on particle $QC$. In a reference frame in which the 
source particle $SA$ is at rest, the test particle $QC$ is at a distance $R = KQ = MA = OA - OM = OA - OQ \cos(\theta) = \rho - a \cos(\theta)$.
If the particle at $Q$ was at rest relative to the source particle at $S$, then it would feel an advanced four-force density with a real component
$\frac{-1}{s_o} \frac{Q q}{2 R^2}$ along $\overrightarrow{KQ}$. But when the particle $QC$ is in hyperbolic motion, it will feel 
an advanced four-force density with a real component $\frac{-1}{s_o} \frac{Q q}{2 R^2} \frac{QC}{SA}$ along $\overrightarrow{KQ}$.
The real component of the advanced four-force with which particle $SA$ is acting on particle $QC$, along $\overrightarrow{KQ}$, in the worldline string model, is 
\begin{equation}
\frac{-1}{s_o} \frac{Q q}{2 R^2} \frac{QC}{SA} QC.
\label{eq:reaction}
\end{equation}
Similar equations describe the imaginary components of the four-forces.
In magnitude the four-force of action (\ref{eq:action}) will be equal to the four-force of reaction (\ref{eq:reaction}) only when 
\begin{equation}
\frac{SA}{r^2} = \frac{QC^2}{R^2 SA^2},
\end{equation}
which reduces to the condition $R = r \frac{QC}{SA} = r \frac{a}{\rho}$. This last equation is easily derived from equation (29) of \cite{galeriuHYP},
as a result of the substitution $y = z = 0$. 

\begin{figure}[h!]
\begin{center}
\includegraphics[height=212pt]{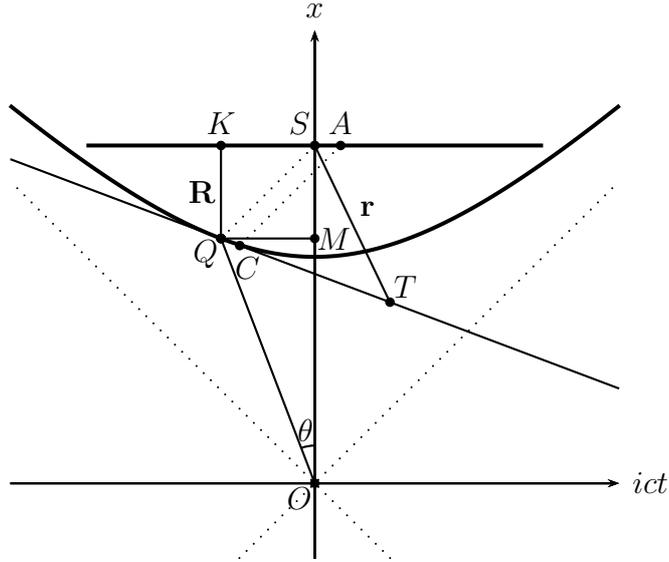}
\caption{A test particle $SA$, at rest in a reference frame in which the source particle $QC$ is in hyperbolic motion, 
and in which the test particle and the center of the hyperbola $O$ are simultaneous, feels a retarded four-force (the action) from the particle $QC$. 
The particle $QC$ feels an advanced four-force (the reaction) from the particle $SA$. The four-force of action and the four-force of reaction are 
equal in magnitude, but opposite in direction.}
\label{fig:action_reaction}
\end{center}
\end{figure}

\section{Three Dimensional Motion}

\subsection{General Formulas}

Consider two points, $\mathbf{X}_1$ (on the worldline of the source particle) 
and $\mathbf{X}_2$ (on the worldline of the test particle) connected by a light signal. Consider two other points infinitely close to the first two points, $\mathbf{X}_1 + \mathbf{\delta X}_1$ (on the worldline of the source particle) 
and $\mathbf{X}_2 + \mathbf{\delta X}_2$ (on the worldline of the test particle), also connected by a light signal. 
We know that 
\begin{equation}
(\mathbf{X}_1 - \mathbf{X}_2) \cdot (\mathbf{X}_1 - \mathbf{X}_2) = 0,
\end{equation}
and that
\begin{equation}
(\mathbf{X}_1 + \delta \mathbf{X}_1 - \mathbf{X}_2 - \delta \mathbf{X}_2) \cdot (\mathbf{X}_1 + \delta \mathbf{X}_1 - \mathbf{X}_2 - \delta \mathbf{X}_2) = 0.
\end{equation}
By rearanging terms in the last equation we have
\begin{equation} 
\left((\mathbf{X}_1 - \mathbf{X}_2) + (\delta \mathbf{X}_1 - \delta \mathbf{X}_2)\right) \cdot 
\left((\mathbf{X}_1 - \mathbf{X}_2) + (\delta \mathbf{X}_1 - \delta \mathbf{X}_2)\right) = 0.
\end{equation}
Expanding this dot product we obtain
\begin{equation}
(\mathbf{X}_1 - \mathbf{X}_2) \cdot (\mathbf{X}_1 - \mathbf{X}_2) 
+ 2 (\mathbf{X}_1 - \mathbf{X}_2) \cdot (\delta \mathbf{X}_1 - \delta \mathbf{X}_2)
+ (\delta \mathbf{X}_1 - \delta \mathbf{X}_2) \cdot (\delta \mathbf{X}_1 - \delta \mathbf{X}_2) = 0.
\end{equation}
To first order in the infinitesimals we have
\begin{equation}
(\mathbf{X}_1 - \mathbf{X}_2) \cdot (\delta \mathbf{X}_1 - \delta \mathbf{X}_2) = 0,
\end{equation}
an equation that we can also write as
\begin{equation}
(\mathbf{X}_1 - \mathbf{X}_2) \cdot \delta \mathbf{X}_1 = (\mathbf{X}_1 - \mathbf{X}_2) \cdot \delta \mathbf{X}_2.
\label{eq:fokker}
\end{equation}

The equation (\ref{eq:fokker}) is exactly the form in which Fokker \cite{fokker} implements in his variational method the assumption that the 
interaction takes place between 
corresponding effective elements ("entsprechended effektiven Elementen") of infinitesimal length. 
These corresponding segments, along the worldlines of the particles, 
have their endpoints connected by light signals.
In Fokker's article, with his notation, the relation (\ref{eq:fokker}) is written as $(R \cdot dy) = (R \cdot dx)$.

With the substitutions $\delta \mathbf{X}_1 = \mathbf{V}_1 \delta \tau_1$ and $\delta \mathbf{X}_2 = \mathbf{V}_2 \delta \tau_2$, 
the ratio of the two infinitesimal segments connected by light signals at both ends is \cite{lamont}:
\begin{equation}
\frac{\delta s_1}{\delta s_2}
= \frac{i c \delta \tau_1}{i c \delta \tau_2} 
= \frac{(\mathbf{X}_1 - \mathbf{X}_2) \cdot \mathbf{V}_2}{(\mathbf{X}_1 - \mathbf{X}_2) \cdot \mathbf{V}_1}.
\label{eq:lamont}
\end{equation}

\subsection{Source Charge at Rest}

When the source charge is moving with a uniform velocity we work in an inertial reference frame in which the source particle is at rest, $\vec{v_1} = 0$.
In this special reference frame we have:
\begin{equation}
\mathbf{V}_1 = (\vec{0}, i c), \ \ \ {\rm (retarded\ or\ advanced)}
\label{eq:rest_v1}
\end{equation}
\begin{equation}
\mathbf{V}_2 = (\gamma_2 \vec{v_2}, i \gamma_2 c),
\label{eq:rest_v2}
\end{equation}
\begin{equation}
\mathbf{X}_2 - \mathbf{X}_{1 ret} = ( \vec{R}, i R ), \ \ \ {\rm (retarded)}
\label{eq:rest_ret}
\end{equation}
\begin{equation}
\mathbf{X}_2 - \mathbf{X}_{1 adv} = ( \vec{R}, -i R ), \ \ \ {\rm (advanced)}
\label{eq:rest_adv}
\end{equation}
and equation (\ref{eq:lamont}) becomes
\begin{equation}
\frac{\delta s_{1 ret}}{\delta s_2}
= \frac{( \vec{R}, i R ) \cdot (\gamma_2 \vec{v_2}, i \gamma_2 c)}{( \vec{R}, i R ) \cdot (\vec{0}, i c)}
= \frac{\gamma_2 \vec{R} \cdot \vec{v_2} - \gamma_2 R c}{- R c}
= \gamma_2 \left( 1 - \frac{v_{2 rad}}{c} \right),
\label{eq:galeriu1}
\end{equation}
\begin{equation}
\frac{\delta s_{1 adv}}{\delta s_2}
= \frac{( \vec{R}, -i R ) \cdot (\gamma_2 \vec{v_2}, i \gamma_2 c)}{( \vec{R}, -i R ) \cdot (\vec{0}, i c)}
= \frac{\gamma_2 \vec{R} \cdot \vec{v_2} + \gamma_2 R c}{R c}
= \gamma_2 \left( 1 + \frac{v_{2 rad}}{c} \right),
\label{eq:galeriu2}
\end{equation}
where the radial velocity of the test particle is $v_{2 rad} = \frac{\vec{R} \cdot \vec{v_2}}{R}$ and $\gamma_2 = \frac{1}{\sqrt{1-\frac{v_2^2}{c^2}}}$.
A geometrical derivation of equations (\ref{eq:galeriu1}) and (\ref{eq:galeriu2}) was provided in \cite{galeriuAAAD}. 
In the one dimensional case $v_{2 rad} = v_2 \equiv v$
and we recover the results from the previous section. In the three dimensional case, however, from the calculation of the four-force we
obtain a retarded component
\begin{equation}
\frac{Q q}{2 R^2} \frac{\sqrt{1-\frac{v_2^2}{c^2}}}{1-\frac{v_{2 rad}}{c}} (\frac{\vec{R}}{R}, i),
\label{eq:ret4force}
\end{equation}
and an advanced component
\begin{equation}
\frac{Q q}{2 R^2} \frac{\sqrt{1-\frac{v_2^2}{c^2}}}{1+\frac{v_{2 rad}}{c}} (\frac{\vec{R}}{R}, - i).
\label{eq:adv4force}
\end{equation}
The total four-force has a real component in the direction of $\vec{R}$
\begin{equation}
\frac{Q q}{2 R^2} \frac{\sqrt{1-\frac{v_2^2}{c^2}}}{1-\frac{v_{2 rad}}{c}} 
+ \frac{Q q}{2 R^2} \frac{\sqrt{1-\frac{v_2^2}{c^2}}}{1+\frac{v_{2 rad}}{c}}
= \frac{Q q}{R^2} \frac{1}{\sqrt{1-\frac{v_2^2}{c^2}}} \frac{1-\frac{v_2^2}{c^2}}{1-\frac{v_{2 rad}^2}{c^2}},
\label{eq:real4force}
\end{equation}
and an imaginary component
\begin{equation}
i \frac{Q q}{2 R^2} \frac{\sqrt{1-\frac{v_2^2}{c^2}}}{1-\frac{v_{2 rad}}{c}} 
- i \frac{Q q}{2 R^2} \frac{\sqrt{1-\frac{v_2^2}{c^2}}}{1+\frac{v_{2 rad}}{c}}
= i \frac{Q q}{R^2} \frac{v_{2 rad}}{c} \frac{1}{\sqrt{1-\frac{v_2^2}{c^2}}} \frac{1-\frac{v_2^2}{c^2}}{1-\frac{v_{2 rad}^2}{c^2}},
\label{eq:imag4force}
\end{equation} 
which differ from the accepted expressions by a factor of $\frac{1 - v_2^2 / c^2}{1 - v_{2 rad}^2 / c^2}$. 
We can have a perfect match with the textbook expressions, but only when the velocity is radial, or zero. We also notice that 
the total four-force (\ref{eq:real4force}) - (\ref{eq:imag4force}) is orthogonal to the four-velocity (\ref{eq:rest_v2}) of the test particle.

\subsection{Source Charge in Hyperbolic Motion}

When the source charge is moving along the $x$ axis with a uniform acceleration 
we work in a reference frame in which the test point and the center of the hyperbola are synchronous, as shown in Figure \ref{fig:combined}. The trajectory of the source particle
is given by $(a \cos(\psi - \varphi), 0, 0, a \sin(\psi - \varphi))$, and its four-velocity
is given by $(- i c \sin(\psi - \varphi), 0, 0, i c \cos(\psi - \varphi))$,
where $\psi - \varphi = - \theta$, as discussed in \cite{galeriuHYP}. 
Angle $\theta$ is the solution to equation (4) of \cite{galeriuHYP}. That equation in fact
has two solutions: a positive angle $\theta$ for the position of the retarded source, and 
a negative angle of the same magnitude $- \theta$ for the position of the advanced source. 
In the special reference frame of Figure \ref{fig:combined} we have:
\begin{equation}
\mathbf{V}_{1 ret} = (i c \sin(\theta), 0, 0, i c \cos(\theta)) = (- \frac{r c}{\rho}, 0, 0, i c \cos(\theta)), \ \ \ {\rm (retarded)}
\label{eq:hyp_v1ret}
\end{equation}
\begin{equation}
\mathbf{V}_{1 adv} = (- i c \sin(\theta), 0, 0, i c \cos(\theta)) = (\frac{r c}{\rho}, 0, 0, i c \cos(\theta)), \ \ \ {\rm (advanced)}
\label{eq:hyp_v1adv}
\end{equation}
\begin{equation}
\mathbf{V}_2 = (\gamma_2 \vec{v_2}, i \gamma_2 c),
\label{eq:hyp_v2}
\end{equation}
\begin{equation}
\mathbf{X}_2 - \mathbf{X}_{1 ret} = ( \rho - a \cos(\theta), y, z, i \frac{r a}{\rho} ) \equiv(\vec{R}, i R), \ \ \ {\rm (retarded)}
\label{eq:hyp_ret}
\end{equation}
\begin{equation}
\mathbf{X}_2 - \mathbf{X}_{1 adv} = ( \rho - a \cos(\theta), y, z, - i \frac{r a}{\rho} ) \equiv(\vec{R}, - i R), \ \ \ {\rm (advanced)}
\label{eq:hyp_adv}
\end{equation}
where we have used the fact that $\overrightarrow{MP} = (\rho - a \cos(\theta), y, z) \equiv \vec{R}$ and 
that $MP = \frac{r a}{\rho} \equiv R$, results derived on page 371 of \cite{galeriuHYP}.
For the derivation of (\ref{eq:hyp_v1ret}) and (\ref{eq:hyp_v1adv}) we have also used the fact 
that $i \sin(\theta) = - \frac{r}{\rho}$, which is a consequence of the equation
$r = - i \rho \sin(\theta)$ derived on page 369 of \cite{galeriuHYP}.
Equation (\ref{eq:lamont}) becomes
\begin{multline}
\frac{\delta s_{1 ret}}{\delta s_2}
= \frac{( \vec{R}, i R ) \cdot (\gamma_2 \vec{v_2}, i \gamma_2 c)}
{( \rho - a \cos(\theta), y, z, i \frac{r a}{\rho} ) \cdot (- \frac{r c}{\rho}, 0, 0, i c \cos(\theta))} \\
= \frac{\gamma_2 \vec{R} \cdot \vec{v_2} - \gamma_2 R c}{- r c} 
= \gamma_2 \left( 1 - \frac{v_{2 rad}}{c} \right) \frac{a}{\rho},
\label{eq:galeriu3}
\end{multline}
\begin{multline}
\frac{\delta s_{1 adv}}{\delta s_2}
= \frac{( \vec{R}, -i R ) \cdot (\gamma_2 \vec{v_2}, i \gamma_2 c)}
{( \rho - a \cos(\theta), y, z, - i \frac{r a}{\rho} ) \cdot (\frac{r c}{\rho}, 0, 0, i c \cos(\theta))} \\
= \frac{\gamma_2 \vec{R} \cdot \vec{v_2} + \gamma_2 R c}{r c}
= \gamma_2 \left( 1 + \frac{v_{2 rad}}{c} \right) \frac{a}{\rho},
\label{eq:galeriu4}
\end{multline}
where the radial velocity of the test particle is $v_{2 rad} = \frac{\vec{R} \cdot \vec{v_2}}{R}$ and $\gamma_2 = \frac{1}{\sqrt{1-\frac{v_2^2}{c^2}}}$.
In the one dimensional case $v_{2 rad} = v_2 \equiv v$
and we recover the results from the previous section. In the three dimensional case, however, from the calculation of the four-force we
obtain a retarded component
\begin{equation}
\frac{Q q}{2 r^2} \frac{\sqrt{1-\frac{v_2^2}{c^2}}}{1-\frac{v_{2 rad}}{c}} (\frac{\vec{R}}{R}, i),
\label{eq:ret4forcehyp}
\end{equation}
and an advanced component
\begin{equation}
\frac{Q q}{2 r^2} \frac{\sqrt{1-\frac{v_2^2}{c^2}}}{1+\frac{v_{2 rad}}{c}} (\frac{\vec{R}}{R}, - i).
\label{eq:adv4forcehyp}
\end{equation}
The factor $a / \rho$ disappears when we transition from the reference frame in which the source charge is at rest to the reference frame of Figure 1.
The total four-force has a real component in the direction of $\overrightarrow{MP}$
\begin{equation}
\frac{Q q}{2 r^2} \frac{\sqrt{1-\frac{v_2^2}{c^2}}}{1-\frac{v_{2 rad}}{c}} 
+ \frac{Q q}{2 r^2} \frac{\sqrt{1-\frac{v_2^2}{c^2}}}{1+\frac{v_{2 rad}}{c}}
= \frac{Q q}{r^2} \frac{1}{\sqrt{1-\frac{v_2^2}{c^2}}} \frac{1-\frac{v_2^2}{c^2}}{1-\frac{v_{2 rad}^2}{c^2}},
\label{eq:real4forcehyp}
\end{equation}
and an imaginary component
\begin{equation}
i \frac{Q q}{2 r^2} \frac{\sqrt{1-\frac{v_2^2}{c^2}}}{1-\frac{v_{2 rad}}{c}} 
- i \frac{Q q}{2 r^2} \frac{\sqrt{1-\frac{v_2^2}{c^2}}}{1+\frac{v_{2 rad}}{c}}
= i \frac{Q q}{r^2} \frac{v_{2 rad}}{c} \frac{1}{\sqrt{1-\frac{v_2^2}{c^2}}} \frac{1-\frac{v_2^2}{c^2}}{1-\frac{v_{2 rad}^2}{c^2}},
\label{eq:imag4forcehyp}
\end{equation} 
which differ from the accepted expressions by a factor of $\frac{1 - v_2^2 / c^2}{1 - v_{2 rad}^2 / c^2}$. 
We can have a perfect match with the textbook expressions, but only when the velocity is radial, or zero. We also notice that 
the total four-force (\ref{eq:real4forcehyp}) - (\ref{eq:imag4forcehyp}) is orthogonal to the four-velocity (\ref{eq:hyp_v2}) of the test particle.

\subsection{Discussion (Part 1)}

What is the origin of this unexpected factor of $\frac{1 - v^2 / c^2}{1 - v_{rad}^2 / c^2}$?
Do we have to modify our theory in order to make this factor disappear, or shall we leave it in place?
Does the average contribution of this factor vanish for periodic motions, 
or for collisions in which the interaction is limited to a restricted spacetime region,
or maybe for any kind of motion?

When investigating the most general expression of the electrodynamic interaction,
Tetrode \cite{tetrode} has determined that, based on the theory of invariants, there are only
four scalar products that are needed: 
$(\mathbf{X}_2 - \mathbf{X}_1)^2$, 
$(\mathbf{X}_2 - \mathbf{X}_1) \cdot \mathbf{V}_1$,
$(\mathbf{X}_2 - \mathbf{X}_1) \cdot \mathbf{V}_2$, 
and $\mathbf{V}_1 \cdot \mathbf{V}_2$.

With the help of Tetrode's scalar products we will show that the four-force obtained 
from our geometrical model is very closely related to the "electrostatic" ("elektrostatischen") four-force obtained 
by Fokker.

In the expression of the electrodynamic four-force of Fokker \cite{fokker}, the term responsible for the retarded "electrostatic" interaction, 
multiplied by the infinitesimal of the proper time of the test particle,
exactly as written in Fokker's article with his notation notation, is:
\begin{equation}
\frac{e_1\ e_2}{8\pi c} R \frac{(dx \cdot dy)^2}{(R \cdot dx) (R \cdot dy)^2}.
\label{eq:fokker_ret}
\end{equation}
Foker uses four-vectors of the type $(c t, x, y, z)$ and scalar products with the metric (+, -, -, -).
We are using four vectors of the type $(x, y, z, i c t)$ and scalar products with the metric (+, +, +, +), 
which is equivalent to four-vectors of the type $(x, y, z, c t)$ and scalar products with the metric (+, +, +, -).
The two different metrics produce scalar products that differ by a sign. 
Another thing to notice is that Fokker uses rational electrostatic units, while we are using Gaussian units, and as a result
we will not have a factor of $4 \pi$ in the denominator. With our metric and with Gaussian units the expression (\ref{eq:fokker_ret}) becomes:
\begin{equation}
- \frac{e_1\ e_2}{2 c} R \frac{(dx \cdot dy)^2}{(R \cdot dx) (R \cdot dy)^2}.
\label{eq:fokker_ret2}
\end{equation}

In our notation $dx = \delta \mathbf{X}_2 = \mathbf{V}_2 \delta \tau_2$, $dy = \delta \mathbf{X}_{1 ret} = \mathbf{V}_{1 ret} \delta \tau_{1 ret}$, 
$R = x - y = \mathbf{X}_2 - \mathbf{X}_{1 ret}$, $e_1 = Q$, and $e_2 = q$. The expression (\ref{eq:fokker_ret2}) becomes:
\begin{equation}
- \frac{Q q}{2 c} (\mathbf{X}_2 - \mathbf{X}_{1 ret}) 
\frac{(\mathbf{V}_2 \cdot \mathbf{V}_{1 ret})^2}{((\mathbf{X}_2 - \mathbf{X}_{1 ret}) \cdot \mathbf{V}_2) 
((\mathbf{X}_2 - \mathbf{X}_{1 ret}) \cdot \mathbf{V}_{1 ret})^2}\delta \tau_2.
\label{eq:fokker_ret3}
\end{equation}

We will show that the retarded four-force obtained 
from our geometrical model is very closely related to the retarded "electrostatic" four-force of Fokker (\ref{eq:fokker_ret3}).
In the numerator of (\ref{eq:fokker_ret3}) we replace $(\mathbf{V}_2 \cdot \mathbf{V}_{1 ret})^2$ with 
$(\mathbf{V}_2 \cdot \mathbf{V}_2)(\mathbf{V}_{1 ret} \cdot \mathbf{V}_{1 ret}) = c^4$, thus obtaining:
\begin{equation}
- \frac{Q q}{2 c} (\mathbf{X}_2 - \mathbf{X}_{1 ret}) 
\frac{c^4}{((\mathbf{X}_2 - \mathbf{X}_{1 ret}) \cdot \mathbf{V}_2) 
((\mathbf{X}_2 - \mathbf{X}_{1 ret}) \cdot \mathbf{V}_{1 ret})^2}\delta \tau_2.
\label{eq:galeriu_ret}
\end{equation}

In the expression of the electrodynamic four-force of Fokker \cite{fokker}, the term responsible for the advanced "electrostatic" interaction, 
multiplied by the infinitesimal of the proper time of the test particle,
exactly as written in Fokker's article with his notation notation, is:
\begin{equation}
\frac{e_1\ e_2}{8\pi c} (-R') \frac{(dx \cdot dy')^2}{(R' \cdot dx) (R' \cdot dy')^2}.
\label{eq:fokker_adv}
\end{equation}
With our metric and with Gaussian units the expression (\ref{eq:fokker_ret}) becomes:
\begin{equation}
- \frac{e_1\ e_2}{2 c} (-R') \frac{(dx \cdot dy')^2}{(R' \cdot dx) (R' \cdot dy')^2}.
\label{eq:fokker_adv2}
\end{equation}

In our notation $dx = \delta \mathbf{X}_2 = \mathbf{V}_2 \delta \tau_2$, $dy' = \delta \mathbf{X}_{1 adv} = \mathbf{V}_{1 adv} \delta \tau_{1 adv}$, 
$-R' = x - y' = \mathbf{X}_2 - \mathbf{X}_{1 adv}$, $e_1 = Q$, and $e_2 = q$. The expression (\ref{eq:fokker_adv2}) becomes:
\begin{equation}
\frac{Q q}{2 c} (\mathbf{X}_2 - \mathbf{X}_{1 adv}) 
\frac{(\mathbf{V}_2 \cdot \mathbf{V}_{1 adv})^2}{(-(\mathbf{X}_2 - \mathbf{X}_{1 adv}) \cdot \mathbf{V}_2) 
(-(\mathbf{X}_2 - \mathbf{X}_{1 adv}) \cdot \mathbf{V}_{1 adv})^2}\delta \tau_2.
\label{eq:fokker_adv3}
\end{equation}

We will show that the advanced four-force obtained 
from our geometrical model is very closely related to the advanced "electrostatic" four-force of Fokker (\ref{eq:fokker_adv3}).
In the numerator of (\ref{eq:fokker_adv3}) we replace $(\mathbf{V}_2 \cdot \mathbf{V}_{1 adv})^2$ with 
$(\mathbf{V}_2 \cdot \mathbf{V}_2)(\mathbf{V}_{1 adv} \cdot \mathbf{V}_{1 adv}) = c^4$, thus obtaining:

\begin{equation}
\frac{Q q}{2 c} (\mathbf{X}_2 - \mathbf{X}_{1 adv}) 
\frac{c^4}{(-(\mathbf{X}_2 - \mathbf{X}_{1 adv}) \cdot \mathbf{V}_2) 
(-(\mathbf{X}_2 - \mathbf{X}_{1 adv}) \cdot \mathbf{V}_{1 adv})^2}\delta \tau_2.
\label{eq:galeriu_adv}
\end{equation}

\subsection{The Four-Force due to a Source Charge at Rest}

When the source particle is at rest, using the equations (\ref{eq:rest_v1}), (\ref{eq:rest_v2}), (\ref{eq:rest_ret}),
(\ref{eq:rest_adv}), the scalar products of interest are:
\begin {equation}
(\mathbf{X}_2 - \mathbf{X}_{1 ret}) \cdot \mathbf{V}_1 = - R\ c,\ \ \ {\rm (retarded)}
\label{eq:sp1}
\end{equation}
\begin {equation}
(\mathbf{X}_2 - \mathbf{X}_{1 adv}) \cdot \mathbf{V}_1 = R\ c,\ \ \ {\rm (advanced)}
\label{eq:sp2}
\end{equation}
\begin {equation}
(\mathbf{X}_2 - \mathbf{X}_{1 ret}) \cdot \mathbf{V}_2 
= \gamma_2 \vec{R} \cdot \vec{v_2} - \gamma_2 R c 
= \gamma_2 R (v_{2 rad} - c),\ \ \ {\rm (retarded)}
\label{eq:sp3}
\end{equation}
\begin {equation}
(\mathbf{X}_2 - \mathbf{X}_{1 adv}) \cdot \mathbf{V}_2 
= \gamma_2 \vec{R} \cdot \vec{v_2} + \gamma_2 R c 
= \gamma_2 R (v_{2 rad} + c),\ \ \ {\rm (advanced)}
\label{eq:sp4}
\end{equation}
\begin {equation}
\mathbf{V}_1 \cdot \mathbf{V}_2 = - \gamma_2 c^2.\ \ \ {\rm (retarded\ or\ advanced)}
\label{eq:sp5}
\end{equation}

With the four-vectors (\ref{eq:rest_v1}), (\ref{eq:rest_v2}), (\ref{eq:rest_ret}) 
and the scalar products (\ref{eq:sp1}), (\ref{eq:sp3}), (\ref{eq:sp5}) 
the expression (\ref{eq:fokker_ret3}) becomes:
\begin{equation}
- \frac{Q q}{2 c} ( \vec{R}, i R ) 
\frac{(- \gamma_2 c^2)^2}{\gamma_2 R (v_{2 rad} - c) (- R\ c)^2}\delta \tau_2
= \frac{Q q}{2} \frac{( \vec{R}, i R )}{R^3} 
\frac{\gamma_2\ \delta \tau_2}{(1 - v_{2 rad}/c)},
\label{eq:fokker_ret4}
\end{equation}
and the expression (\ref{eq:galeriu_ret}) becomes:
\begin{equation}
- \frac{Q q}{2 c} ( \vec{R}, i R ) 
\frac{c^4}{\gamma_2 R (v_{2 rad} - c) (- R\ c)^2}\delta \tau_2
= \frac{Q q}{2} \frac{( \vec{R}, i R )}{R^3} 
\frac{\delta \tau_2}{\gamma_2 (1 - v_{2 rad}/c)},
\label{eq:galeriu_ret4}
\end{equation}
which is the same as the four-force (\ref{eq:ret4force}) multiplied by $\delta \tau_2$.

With the four-vectors (\ref{eq:rest_v1}), (\ref{eq:rest_v2}), (\ref{eq:rest_adv}) 
and the scalar products (\ref{eq:sp2}), (\ref{eq:sp4}), (\ref{eq:sp5}) 
the expression (\ref{eq:fokker_adv3}) becomes:
\begin{equation}
- \frac{Q q}{2 c} ( \vec{R}, - i R ) 
\frac{(- \gamma_2 c^2)^2}{-\gamma_2 R (v_{2 rad} + c) (R\ c)^2}\delta \tau_2
= \frac{Q q}{2} \frac{( \vec{R}, - i R )}{R^3} 
\frac{\gamma_2\ \delta \tau_2}{(1 + v_{2 rad}/c)},
\label{eq:fokker_adv4}
\end{equation}
and the expression (\ref{eq:galeriu_adv}) becomes:
\begin{equation}
- \frac{Q q}{2 c} ( \vec{R}, - i R ) 
\frac{c^4}{-\gamma_2 R (v_{2 rad} + c) (R\ c)^2}\delta \tau_2
= \frac{Q q}{2} \frac{( \vec{R}, - i R )}{R^3} 
\frac{\delta \tau_2}{\gamma_2 (1 + v_{2 rad}/c)},
\label{eq:galeriu_adv4}
\end{equation}
which is the same as the four-force (\ref{eq:adv4force}) multiplied by $\delta \tau_2$.
We note at this time that the extra factor of $\gamma_2^2$ in Fokker's expressions is still not enough to cancel 
completely the factor of $\frac{1 - v_2^2 / c^2}{1 - v_{2 rad}^2 / c^2}$ in order to reproduce the textbook values.

\subsection{The Four-Force due to a Source Charge in Hyperbolic Motion}

When the source particle is in hyperbolic motion, using the equations (\ref{eq:hyp_v1ret}), (\ref{eq:hyp_v1adv}), 
(\ref{eq:hyp_v2}), (\ref{eq:hyp_ret}), (\ref{eq:hyp_adv}), the scalar products of interest are:
\begin {equation}
(\mathbf{X}_2 - \mathbf{X}_{1 ret}) \cdot \mathbf{V}_{1 ret} = - r\ c,\ \ \ {\rm (retarded)}
\label{eq:sp1hyp}
\end{equation}
\begin {equation}
(\mathbf{X}_2 - \mathbf{X}_{1 adv}) \cdot \mathbf{V}_{1 adv} = r\ c,\ \ \ {\rm (advanced)}
\label{eq:sp2hyp}
\end{equation}
\begin {equation}
(\mathbf{X}_2 - \mathbf{X}_{1 ret}) \cdot \mathbf{V}_2 
= \gamma_2 \vec{R} \cdot \vec{v_2} - \gamma_2 R c 
= \gamma_2 R (v_{2 rad} - c),\ \ \ {\rm (retarded)}
\label{eq:sp3hyp}
\end{equation}
\begin {equation}
(\mathbf{X}_2 - \mathbf{X}_{1 adv}) \cdot \mathbf{V}_2 
= \gamma_2 \vec{R} \cdot \vec{v_2} + \gamma_2 R c 
= \gamma_2 R (v_{2 rad} + c),\ \ \ {\rm (advanced)}
\label{eq:sp4hyp}
\end{equation}
\begin {equation}
\mathbf{V}_{1 ret} \cdot \mathbf{V}_2 
= - \frac{r c}{\rho} \gamma_2 v_{2x} - \gamma_2 c^2 \cos(\theta)
= - \gamma_2 c^2 ( \cos(\theta) + \frac{r v_{2x}}{\rho c} ),\ \ \ {\rm (retarded)}
\label{eq:sp5hyp}
\end{equation}
\begin {equation}
\mathbf{V}_{1 adv} \cdot \mathbf{V}_2 
= \frac{r c}{\rho} \gamma_2 v_{2x} - \gamma_2 c^2 \cos(\theta)
= - \gamma_2 c^2 ( \cos(\theta) - \frac{r v_{2x}}{\rho c} ).\ \ \ {\rm (advanced)}
\label{eq:sp6hyp}
\end{equation}

With the four-vectors (\ref{eq:hyp_v1ret}), (\ref{eq:hyp_v2}), (\ref{eq:hyp_ret}) 
and the scalar products (\ref{eq:sp1hyp}), (\ref{eq:sp3hyp}), (\ref{eq:sp5hyp}) 
the expression (\ref{eq:fokker_ret3}) becomes:
\begin{equation}
- \frac{Q q}{2 c} ( \vec{R}, i R ) 
\frac{(- \gamma_2 c^2)^2 ( \cos(\theta) + \frac{r v_{2x}}{\rho c} )^2}{\gamma_2 R (v_{2 rad} - c) (- r\ c)^2}\delta \tau_2
= \frac{Q q}{2} \frac{( \vec{R}, i R )}{R\ r^2} 
\frac{\gamma_2\ ( \cos(\theta) + \frac{r v_{2x}}{\rho c} )^2 \delta \tau_2}{(1 - v_{2 rad}/c)},
\label{eq:fokker_ret5}
\end{equation}
and the expression (\ref{eq:galeriu_ret}) becomes:
\begin{equation}
- \frac{Q q}{2 c} ( \vec{R}, i R ) 
\frac{c^4}{\gamma_2 R (v_{2 rad} - c) (- r\ c)^2}\delta \tau_2
= \frac{Q q}{2} \frac{( \vec{R}, i R )}{R\ r^2} 
\frac{\delta \tau_2}{\gamma_2 (1 - v_{2 rad}/c)},
\label{eq:galeriu_ret5}
\end{equation}
which is the same as the four-force (\ref{eq:ret4forcehyp}) multiplied by $\delta \tau_2$.

With the four-vectors (\ref{eq:hyp_v1adv}), (\ref{eq:hyp_v2}), (\ref{eq:hyp_adv}) 
and the scalar products (\ref{eq:sp2hyp}), (\ref{eq:sp4hyp}), (\ref{eq:sp6hyp}) 
the expression (\ref{eq:fokker_adv3}) becomes:
\begin{equation}
- \frac{Q q}{2 c} ( \vec{R}, - i R ) 
\frac{(- \gamma_2 c^2)^2 ( \cos(\theta) - \frac{r v_{2x}}{\rho c} )^2}{-\gamma_2 R (v_{2 rad} + c) (r\ c)^2}\delta \tau_2
= \frac{Q q}{2} \frac{( \vec{R}, - i R )}{R\ r^2} 
\frac{\gamma_2\ ( \cos(\theta) - \frac{r v_{2x}}{\rho c} )^2 \delta \tau_2}{(1 + v_{2 rad}/c)},
\label{eq:fokker_adv5}
\end{equation}
and the expression (\ref{eq:galeriu_adv}) becomes:
\begin{equation}
- \frac{Q q}{2 c} ( \vec{R}, - i R ) 
\frac{c^4}{-\gamma_2 R (v_{2 rad} + c) (r\ c)^2}\delta \tau_2
= \frac{Q q}{2} \frac{( \vec{R}, - i R )}{R\ r^2} 
\frac{\delta \tau_2}{\gamma_2 (1 + v_{2 rad}/c)},
\label{eq:galeriu_adv5}
\end{equation}
which is the same as the four-force (\ref{eq:adv4forcehyp}) multiplied by $\delta \tau_2$.

\subsection{The Law of Action and Reaction}

The relativistic invariant expressions (\ref{eq:fokker_ret2}) and (\ref{eq:fokker_adv2}) of Fokker
are the starting point in a discussion about conservation laws.
When using the material point particle model, the expressions (\ref{eq:fokker_ret2}) and (\ref{eq:fokker_adv2}) are interpreted as the product of
the four-force with the proper time interval during which the four-force acts, 
thus giving the change in the momentum-energy four vector of the particle. The conservation of the
total momentum-energy is then easily demonstrated \cite{fokker}. 
When using the worldline string model, the expressions (\ref{eq:fokker_ret2}) and (\ref{eq:fokker_adv2}) are interpreted as the product of
the linear four-force density with the infinitesimal length of the worldline segment on which the four-fource density acts,
thus giving the four-force acting on the worldline element. The law of action and reaction is then easily demonstrated.

The proof goes like this. In the formulas (\ref{eq:fokker_ret2}) and (\ref{eq:fokker_adv2}) $x$ gives the place and time of the test particle, 
$y$ gives the place and time of the retarded source particle, and $y'$ gives the place and time of the advanced source particle. 
We know the retarded four-force (\ref{eq:fokker_ret2}) with which the particle at $y$ acts upon the particle at $x$, and we want to find the 
advanced four-force with which the particle at $x$ acts upon the particle at $y$. We can find this advanced four-force with the help of formula 
(\ref{eq:fokker_adv2}), simply by replacing $y'$ with $x$, by replacing $x$ with $y$, and by replacing $R' = y' - x$ with $R = x - y$. 
Because $(R \cdot dx) = (R \cdot dy)$ the expression thus obtained is equal in magnitude to the expression (\ref{eq:fokker_ret2}), 
but opposite in direction, due to an extra minus sign. 
The same reasoning applies to our theory, in which we replace 
$(dx \cdot dy)^2$ with $(dx \cdot dx) (dy \cdot dy)$ 
in the numerator of (\ref{eq:fokker_ret2}) and (\ref{eq:fokker_adv2}).

\subsection{Discussion (Part 2)}

At this point in our research it seems that the $\frac{1 - v^2 / c^2}{1 - v_{rad}^2 / c^2}$ 
factor is an essential part of our theory.
Fokker's theory also exhibits strange extra factors. Fokker himself noticed that
in the expression of his four-force there are "dependencies upon the velocities neglected in the usual derivations"
("bei den \"{u}blichen Ableitungen vernachl\"{a}ssigten Abh\"{a}ngigkeiten von den Geschwindigkeiten").
This surprising message did not survive in the modern physics literature.
What we are currently referring to as Fokker's time symmetric action is in fact a modern expression that involves
the use of a Dirac delta function. Essential features of the original theory, 
like the strange extra factors, or the possibility
of having a four-force that is not orthogonal to the four-velocity, have been lost during the
conversion from the original formalism (involving corresponding infinitesimal segments) 
to the Dirac delta formalism (involving point particles). Ironically, in 1938 Dirac himself \cite{dirac} 
has mentioned Fokker's action in its 1929 original form, without using the Dirac delta function.

\section{Conclusions}

Our geometrical analysis of the electromagnetic interaction between electric charges at rest, in uniform motion, or in hyperbolic motion, has revealed
the fact that the material point particle model is equivalent to a model in which the worldline of a particle is a string in static equilibrium.
The four-force acting on a point particle is equivalent to a linear four-force density acting on an infinitesimal length element along the worldline of the particle.
The retarded four-force density and the advanced four-force density are added together, and then multiplied by the length $s_o$ of the particle
in order to produce the total four-force acting on the particle. 
The time symmetric interaction between the infinitesimal segments is described by some very simple rules:

1) Corresponding infinitesimal segments who interact have their endpoints connected by light signals. 

2) The four-force of action and the four-force of reaction
are equal in magnitude and opposite in direction, along the line connecting the two particles.

3) The distance $R$ in Coulomb's formula and the retarded or advanced linear four-force density acting on the test particle 
are measured in the reference frame in which the source particle is at rest.

4) The length of the source particle $CD$ is kept constant, while the length of the test particle $AB$ is allowed to vary.

5) In the reference frame in which the source particle is at rest, the linear four-force density is equal to the product 
of $- \frac{1}{s_o} \frac{Q q}{2 R^2}$ with $\frac{AB}{CD}$, where $Q$ is the electric charge of the source particle
and $q$ is the electric charge of the test particle.

6) The retarded linear four-force density and the advanced linear four-force density are added together, in order to give the linear four-force density.

7) The linear four-force density is multiplied by the constant length $s_o$ of the test particle, and by $-1$, in order to give the four-force.

From a practical point of view, one could ignore the $- \frac{1}{s_o}$ factor in Step 5, and the corresponding multiplication by $- s_o$ in step 7.
We should keep in mind that we cannot add four-forces that act on worldline segments of different length, but we can always add 
the linear force densities.

In the one dimensional case this geometrical recipe reproduces the classical formula of the electrodynamic four-force. 
In the three dimensional case we get an extra factor of $\frac{1 - v^2 / c^2}{1 - v_{rad}^2 / c^2}$. 
This extra factor, together with the possibility of having a four-force that is not orthogonal to the four-velocity, are very important features of
our theory that should not be ignored, but further investigated.

We will conclude this article with an interesting observation. 
Was Minkowski himself on the verge of discovering the possibility of replacing the interacting material point particles with infinitesimal length elements 
along their worldlines? This is indeed very likely, given the words "Let BC be an infinitely small element of the worldline of F; 
further let B* be the light point of B, C* be the light point of C on the worldline of F* [...]" that he uses when describing his 
first theory of gravitational interaction \cite{minkowski2}.

\section{Acknowledgments}

The author is very much indebted to David~H.~Delphenich for 
translating from German into English the articles by Tetrode and Fokker, upon personal request.

This manuscript is dedicated to the memory of my late father, Dr. Dan C\u{a}lin Constantin Galeriu.

\end{document}